\documentclass[]{aa}

\usepackage{graphicx}
\usepackage{graphics}
\usepackage{epsfig}
\newcommand{\be}{\begin{equation}}
\newcommand{\ee}{\end{equation}}

\begin{document}

\title{Gamma-ray and neutrino emission from misaligned microquasars}

\author{Gustavo E. Romero\inst{1,2,}\thanks{Member of CONICET, Argentina} and Mariana Orellana
\inst{1,2,}\thanks{Fellow of CONICET, Argentina}}

\offprints{M. Orellana\\  \email{morellana@irma.iar.unlp.edu.ar}}

\titlerunning{High-energy emission from microquasars}

\authorrunning{G.E. Romero \& M. Orellana}

\institute{Instituto Argentino de Radioastronom\'{\i}a, C.C.5,
(1894) Villa Elisa, Buenos Aires, Argentina \and Facultad de
Ciencias Astron\'omicas y Geof\'{\i}sicas, Universidad Nacional
 de La Plata, Paseo del Bosque, 1900 La Plata, Argentina}

\date{Received / Accepted}

\abstract{Microquasars are accreting X-ray binary systems with non-thermal radio jets.
In some of these systems the jet is expected to be
strongly misaligned with the perpendicular to the orbital plane. If the donor star is an early-type star, the
jet could collide with the stellar wind producing a standing shock between the compact object and the stellar surface. Relativistic particles injected by the jet can be re-accelerated and isotropized at the colliding region.
If the jet has hadronic content, TeV protons will diffuse into the inner, dense wind leading to gamma-ray and neutrino production from interactions with the matter of the wind. In the case of very powerful jets, the wind pressure can be overbalanced and the jet might impact directly onto the stellar surface. We present estimates of the gamma-ray and neutrino luminosities for different sets of parameters in these scenarios and we briefly discuss the effects of this radiation on the donor star and its detectability with current instruments.
\keywords{X-ray: binaries -- stars -- gamma-rays: theory -- gamma-rays: observations -- neutrinos}}

\maketitle

\section{Introduction}
Microquasars (MQs) are accreting X-ray binary systems that have shown
persistent or episodic non-thermal jet-like ejections (Mirabel \&
Rodr\'{\i}guez 1999). It is believed that a significant fraction of all accreting black hole (BH) and neutron star (NS)
X-ray binaries are likely to exhibit MQ behavior at
some epoch during their lifetime. Although it is not observationally
required, the jet axis and the orbital plane of a given MQ system
are usually assumed to be approximately perpendicular. However,
Maccarone (2002) has shown that the timescale for the BH spin to align with the orbital angular momentum in
binary systems is often longer than the lifetime of the system
itself. Additionally, tidal effects upon the disk can induce a precession which could result in a severe misalignment (e.g. Larwood 1998, Kaufman et al. 2002). Precession and strong inclination of the jet can also be the result of the radiative warping of the inner accretion disk (e.g. Wijers \& Pringle 1999, Ogilvie \& Dubus 2001) or might be due to spin-spin interactions (e.g. Bardeen \& Petterson 1975, Sarazin et al. 1980). Observationally, there exists suggestive evidence for jet misalignments with the normal to the orbital plane in wind-fed accreting BHs. For instance, the jet axis lies at no more than $\sim 35$ degrees from the orbital plane in V4641 Sgr (see Butt et al. 2003 and references therein). Other systems like SS 433, LSI +61 303, and Cygnus X-3 are known to have precessing jets (see Katz 1980, Massi et al. 2004, and Mioduszewski et al. 2001, respectively). Precessing jets have been also suggested in the case of Cygnus X-1 (Romero et al. 2002).

In this paper we will consider the situation of a misaligned MQs with its jet lying close to the
binary plane and being periodically directed straight to the
companion star. Such a situation seems not to be unlikely. Butt et al. (2003), for
instance, have considered the induced nucleosynthesis that should
occur on the stellar surface as a consequence of the jet impact in
such systems. Here, we will show that in the case of an
early-type companion star, the jet, under some conditions, could not
reach  the stellar surface since it is balanced by the strong
stellar wind. However, another interesting feature appears: a
strong shock is formed between the compact object and the star and
protons injected by the jet can be re-accelerated and isotropized
there. The most energetic of these protons then diffuse through
the inner wind producing gamma-rays and neutrinos through
$p+p\longrightarrow \pi^0 + X$ and  $p+p\longrightarrow \pi^{\pm}
+ X$ reaction channels.

On the other hand, very powerful jets can overcome the pressure of the wind
and directly impact on the star. In such a case the jet
energy is dissipated through hadronic and electromagnetic cascades
that develop in the stellar atmosphere. No continuum gamma-ray
source above MeV energies is expected in this case, but a
significant flux of high-energy neutrinos goes through the star.
In what follows we characterize the general physical situation and
estimate the gamma-ray and neutrino output of these systems, under
a variety of conditions. The results will be of interest for
current ground-based TeV gamma-ray telescopes and future
satellites for MeV-GeV gamma-ray astronomy, as well as for
km-scale neutrino detectors.

\section{Jet-wind interaction}

In order to make specific calculations we will consider a
misaligned MQ with a high-mass companion. The jet axis is assumed,
for simplicity, to be parallel to the orbital radius $a$, and
pointing toward the massive star of radius $R_\star$. As in Romero
et al. (2003), we consider an expanding conical jet, i.e. the radius of
the cross-section is $R_{\rm j}(d)=d\,\theta/2$, where $\theta$
is the opening angle of the jet (in radians), $d$ is the distance from the
compact object, and $r$ is the radial coordinate with
origin at the center of the star ($d=a-r$). Assuming adiabatic expansion,
the magnetic field inside the conical jet changes with $d$ as
\be
B_{\rm j}(d)=B_{\rm j}(d_0)\left(\frac{d_0}{d}\right),
\ee
where $d_0$ is the distance from the jet's apex to the compact
object. We will assume $d_0=50\,R_{\rm g}$, being $R_{\rm g}$ the gravitational radius of the accreting star.

In order to determine the kinetic power of the jet we will adopt
the jet-disk coupling hypothesis proposed by Falcke \& Biermann
(1995) and applied with success to AGNs, i.e. the total jet power
scales with the accretion rate as $L_{\rm j}=q_{\rm j}
\dot{M}_{\rm disk} c^2$. We will assume that relativistic protons provide the dominant contribution to the jet power, i.e. $L_{\rm j}\approx \gamma_{\rm bulk} \dot{N}_p m_p c^2$, where $\dot{N}_p$ is the proton injection rate, $m_p$ the proton mass, and $\gamma_{\rm bulk}$ is the bulk Lorentz factor of the flow. The presence of relativistic hadrons has been inferred from iron X-ray line observations for the case of SS433 (e.g.
Kotani et al. 1994, 1996; Migliari et al. 2002). Although direct and clear evidence exists only for this source so far, several models assume a hadronic content for the jets of some MQs (see Romero 2004 and references therein).

The high-mass companion star loses a significant fraction of its
mass through a very strong supersonic wind, and then the mass transfer
occurs mainly through this stellar wind. To ensure that
there is no Roche lobe overflow the parameters of the binary
system must satisfy the condition $R_L>R_\star$, where the Roche
radius can be expressed in good approximation as
\begin{equation}
{R_L}\approx a\left[0.38+0.2 \log(M_\star/M_{\rm co})\right],
\end{equation}
for $0.3\leq (M_\star/M_{\rm co})\leq 20$ (Kopal 1959). Here, $M_\star$ and $M_{\rm co}$ are the masses of the primary massive star and the compact object, respectively.

Typical mass loss rates and terminal wind velocities for
early-type stars are of the order of $10^{-5}$ M$_\odot$ yr$^{-1}$
and 2500 km s$^{-1}$, respectively (Lamers \& Cassinelli 1999).
The matter field will be determined by the continuity equation:
\begin{equation}
\dot{M}_\star=4\pi r^2 \rho (r) v(r),
\end{equation}
where $\rho$ is the density of the wind and $v(r)=v_{\infty} (1-{R_*}/{r})^{\beta}$ is
its velocity; $v_{\infty}$ is the terminal wind velocity, and the
parameter $\beta$ is $\sim 1$ for very massive stars (Lamers \&
Cassinelli 1999). Hence, assuming a gas dominated by protons, the
particle density results:
\be
n(r)=\frac{\dot{M}_\star}{4\pi m_p r^2 v(r)}.
\ee

The pressure exerted  by the stellar wind at a distance $r$ from
the star is then:
\be \label{Pwind}P_{\rm
wind}=\frac{\dot{M}_\star\,v(r)}{4 \pi r^2}, \ee
whereas the
pressure due to the magnetic field of the star is
\be \label{Pmag} P_{\rm mag}=B(r)^2 / 8\pi, \ee
where $B(r)=\sqrt{B_\phi^2+B_r^2}$. The magnetic field
of the star can be represented by the standard magnetic rotor model (Weber
and Davis 1967; White 1985; Lamers and Cassinelli 1999):
$B_\phi/B_r=(v_\star/v_\infty) (1+r/R_\star)$ and $B_r=B_\star
(R_\star/r)^2$, where $v_\star\cong (0.1-0.2)v_\infty$ is the rotational velocity at the
surface of the star (Conti \& Ebbets 1977), and $B_\star$ is the corresponding stellar magnetic field.

If the jet propagates with an opening angle $\theta$, at a distance
$d$ from the compact object the ram pressure of
the jet plasma is
\be \label{Pj}P_{\rm j}=\frac{L_{\rm
j}}{\pi\,c\, \theta^2 d^2}. \ee
As shown by Bednarek \& Protheroe (1997) for the case of AGN's jets, a strong stellar wind can stop the jet flow before it can reach the stellar surface. The interaction region will be formed by a contact discontinuity and two quasi-parallel shocks, one propagating backwards through the jet and the other moving toward the star through the wind.

The position $r_{\rm s}$ where the shock front will form can be obtained
by equating the expression for the jet pressure to the expressions given in Eqs. (\ref{Pwind}) or
(\ref{Pmag}). The location of the double-shock structure will be determined by the specific parameters assumed for the jet and the stellar wind. As an example, we will work out the details of a family of models sharing the parameters listed in Table \ref{t1}. For these values, it happens that $R_L \approx 1.53\,R_\star$ and there is no Roche lobe overflow.

With this set of basic parameters, the ram pressure of the jet can not be balanced by the wind when $q_{\rm j}> 0.006$. Above this value, however, the magnetic pressure can still balance and stop the jet. For $q_{\rm j}=10^{-2}$, a magnetic field of $B_\star\ge 200$ G is necessary to form a terminal shock front for the jet in the wind region. Early-type stars can support surface magnetic fields of even $\sim 10^3$ G as reported by Donati et al. (2002). In order to study systems where the jet is balanced by the wind, we will assume different values for the stellar magnetic field and the jet power. After discussing these systems we will turn to MQs with very strong jets in Section \ref{heavy-jets}.

\begin{table} [h]
\centering \caption{Basic parameters assumed for the model}
\begin{tabular}{lll}
\hline
%\noalign{\smallskip}
Parameter & Symbol  & Value  \\
%\noalign{\smallskip}
\hline
%\noalign{\smallskip}
%Type of jet &  $\delta$ & 1 \\
Mass of the compact object &
$M_{\rm co}$ & 10 $M_{\sun}$\\ Mass of the donor star&M$_\star$ & 46.2
$M_{\sun}$\\ Jet's injection point & $d_0$ & 50 $R_{\rm g}^1$  \\
Opening angle of the jet & $\theta$ &5$^\circ$\\
Bulk Lorentz factor & $\gamma_{bulk}$ & 10 \\
Radius of the companion star & $R_\star$ &15 $R_{\odot}$ \\
Stellar effective temperature& $T_{\rm eff}$& 40000 K
\\ Mass loss rate & $\dot{M}_{\star}$ & $10^{-5}$ $M_{\odot}$ yr$^{-1}$
\\ Terminal wind velocity & $v_{\infty}$ & 2500 km s$^{-1}$\\
Compact object accretion rate & $\dot{M}_{\rm disk}$ & $10^{-8}$
$M_{\odot}$ yr$^{-1}$ \\ Wind velocity index & $\beta$ & 1 \\
Efficiency of the shock  & $\chi$ & 0.1\\
%Jet's expansion index & $n$ & 2 \\
%Jet's Lorentz factor & $\Gamma$ & 5 \\
%Minimum proton energy & ${E'}_p^{\rm min}$ & 10 GeV \\
%Maximum proton energy & ${E'}_p^{\rm max}$ & 100 TeV \\
Orbital radius & $a$ & 3 $R_\star$\\
%\noalign{\smallskip}
\hline \multicolumn{3}{l} {$^1$$R_{\rm g}=GM_{\rm co}/c^2$.}\cr
\end{tabular}
\label{t1}
\end{table}

%All the power of the jet may in principle be
%extracted by the shock and directed to accelerate particles by the
%first order Fermi mechanism.

\section{Particle re-acceleration and diffusion}

The shocks generated by the interaction between the relativistic plasma in the jet  and the stellar wind can accelerate particles by Fermi mechanism.  For acceleration at a parallel shock we can write the acceleration
rate as (e.g. Biermann \& Strittmatter 1987, Protheroe 1998):

\be
\frac{dE}{dt}|_{\rm acc}=\xi Z e c B, \,\,\,{\rm
with}\,\,\xi\simeq0.015 (u/c)^2,\label{acel} \ee
where $u$ is the shock velocity and all values are in cgs units.
%So $\xi\simeq1.5\times
%10^{-4}$ and $\xi\simeq3.75\times 10^{-3}$ respectively for a
%shock velocity  $u=0.1c$ and $u=0.5 c$.

The highest energy gained by the particles in the process is
imposed by the different energy losses (synchrotron emission,
$pp$ interactions and photopion production) or by the available
space in the acceleration region. For the parameters considered in
this paper the dominant energy loss is due to $pp$ interactions with the stellar wind matter that pervades the acceleration region. For
a relativistic proton ($E>1.22$ GeV) this energy loss can be
estimated as in Mannheim \& Schlickeiser (1994):
\begin{equation}
-\frac{d E}{d t}|_{pp}\approx 0.65 c n(r_{\rm s}) \sigma_{pp}(E-m_pc^2),
\end{equation}
where $n(r_{\rm s})$ is the particle density of the wind close to the shock front, and $\sigma_{pp}$ is the inelastic cross section for $pp$ interaction at the energy $E$.
 The maximum energy for the protons then results from

\begin{eqnarray}
E_{p}^{\rm max}&\simeq &{\rm min}\,[\,1.9\times  10^{15}\left(\frac{u}{c}\right)^2\frac{B_{\rm j}(d_o)}{\rm G}\,{\rm eV},\nonumber \\ && 3\times 10^{14}\,\frac{B_{\rm j}(d_o)}{\rm G}\,
{\rm eV}\,].\label{emax}
\end{eqnarray}
We shall adopt $u/c=0.1$ for the sub-relativistic shock and $u/c=0.5$ for a relativistic case. For the former $E_p^{\rm max}$ is determined by the losses, whereas in the second case it is fixed by the size constrain, i.e. the maximum gyroradius for a proton is $\sim (a-R_\star)$.

The resulting proton spectrum from the diffusive acceleration will be a power-law extending from a minimum energy ($E_p^{\rm min}\sim 10$ GeV\footnote{From Table 1 the Lorentz factor of the jet is $\sim 10$ and then those protons at the lower energy tail of the co-moving particle energy distribution (for which the microscopic Lorentz factor is $\sim 1$) are injected at the shock with energies $\sim 10$ GeV.}) up to $E_p^{\rm max}$:
 \be
 N_p(E)= K_\Gamma E^{-\Gamma},\,\, 10\,m_pc^2<E<E_p^{\rm max}. \label{N_p}
 \ee
The index $\Gamma$ is $\sim 2$ for a strong non-relativistic shock and $\sim 1.5$ for a relativistic shock (e.g. Protheroe 1998). 

Some fraction $\chi$ of the energy transported by the jet will be
transferred to the particles accelerated at the shock front. We will assume here an efficiency of $\chi=0.1$
(Blandford \& Ostriker 1978). Then, the constant $K_\Gamma$ in (\ref{N_p}) can be
obtained from
 \be
\chi\, L_{\rm j}= (a-R_\star)^2 c \int^{E_p^{\rm max}}_{10 m_p c^2}
 N_p(E) E\, dE .
 \ee

%Lo que sigue es muy parecido a lo de Torres, por qué no citarlo y listo?
The resulting population of relativistic protons will diffuse through the stellar wind, but not all of them will penetrate up to the densest parts where the probability of $pp$ interactions is higher because of the modulation effect of the wind (Romero \& Torres 2003, Torres et al. 2004). Whether the particles can reach the base of the wind or not is given by the ratio ($\epsilon$) of the
diffusion and convection timescales, which are given by $t_d=3r^2/D$ and
$t_c=3r/v(r)$, respectively. Here $D$ is the diffusion coefficient defined as (Ginzburg \& Syrovatskii 1964):
\be
D \sim \frac{1}{3}\lambda_r c,
\ee where
$\lambda_r$ is the mean-free-path for diffusion in the radial
direction. Only particles for which $\epsilon<1$
will be able to overcome convection and enter the dense wind
region to produce $\gamma$-rays through $pp$ interactions.

The value of the diffusion coefficient is of course not well-known, especially for such a complex and turbulent medium as the wind of a massive star. Following White (1985) we will make use of the approximation given by V\"olk and Forman (1982), assuming that the
mean-free-path for scattering parallel to the magnetic field
direction is $\lambda_\| \sim 10 r_{\rm g} = 10 E/eB$, where
$r_{\rm g}$ is the proton gyro-radius and $E$ the proton energy. The value of $\lambda$ in the
perpendicular direction is shorter: $\lambda_\bot \sim r_g$. The mean-free-path in the radial direction is then (see Torres et al. 2004):
\be
\lambda_r= r_g ( 10 \cos^2 \theta + \sin^2 \theta),
\ee
where
\be
\cos^{-2} \theta = 1+(B_\phi/B_r)^2.
\ee

Through $\epsilon=1$ we find now the minimum energy $E_p^{\rm
min}(r)$ below which the particles are taken away by convection in
the wind (see the explicit expression for $\epsilon$ in the paper by Torres et al. 2004).
In what follows, we will take $E_p^{\rm min}(r_{\rm s})$ as the minimum energy of
a proton accelerated at the shock front to collide with a proton
of the inner wind.

\section{Gamma-ray production} \label{gamma}

We can now proceed to calculate the $\pi^0$ gamma-ray luminosity
from the inner wind. The differential $\gamma$-ray emissivity from
$\pi^0$-decays can be approximated by \be q_{\gamma}(E_{\gamma})=
4 \pi \sigma_{pp}(E_p)
\frac{2Z^{(\Gamma)}_{p\rightarrow\pi^0}}{\Gamma}\;J_p(E_{\gamma})
\eta_{\rm A}\Theta(E_p-E_p^{\rm min}), \label{q_g}\ee at the
energies of interest here. $Z^{(\Gamma)}_{p\rightarrow\pi^0}$
is the so-called spectrum weighted moment, defined as:
\be
Z^{(\Gamma)}_{p\rightarrow\pi^0}=\int^{1}_{0} x^{\Gamma-1}F_{p\pi^0}(x) dx,
\ee
where $x=E_{\pi^0}/E_{p}$ and $F_{p\pi^0}(x)$ is the dimensionless inclusive cross section (Gaisser 1990, p.31) 
. The parameter $\eta_{\rm A}$ reflects the contribution
from different nuclei in the wind (for the standard composition of
an early-type star we will assume $\eta_{\rm A} \sim 1.5$).
$J_p(E_{\gamma})=(c/4\pi) N_p(E_\gamma)$ is the proton flux
distribution evaluated at $E=E_{\gamma}$ (units of protons per
unit time, solid angle, energy-band, and area).

The inelastic cross section for $pp$ interactions at energy $E_p\approx 6\,\xi_{\pi^0} E_{\gamma}/K$, where $K\sim 0.5$ is the inelasticity coefficient and $\xi_{\pi^0}=1.1(E_p/{\rm GeV})^{1/4}$ is the multiplicity for the neutral pions (e.g. Ginzburg \& Syrovatskii, 1964), can be represented above $E_p\approx 10$ GeV by the expression given in Aharonian \& Atoyan (1996). Typically, at 1 TeV, $\sigma_{pp}\sim 34$ mb. Finally, $\Theta(E_p-E_p^{\rm min})$ is a
Heaviside function that takes into account the fact that only
protons with energies higher than $E_p^{\rm min}(r_{\rm s})$ will
diffuse into the densest part of the wind. The spectral
$\gamma$-ray intensity (photons per unit time and energy-band) is:
\be
 I_{\gamma}(E_{\gamma})=\int n(r) q_{\gamma}(E_\gamma) dV, \ee
where $V$ is the interaction volume. The generated luminosity in a given
band ($E_2, \;E_1$) then is
%The spectral $\gamma$-ray intensity
%(photons per unit time and energy-band) is $
%I_{\gamma}(E_{\gamma})=\int n(r) q_{\gamma}(E_\gamma) dV, $ where
%$V$ is the interaction volume.

\be
L_\gamma=\int_{E_1}^{E_2} E_\gamma\, I_\gamma (E_\gamma)\,
dE_\gamma \ee (see, e.g., Romero et al. 2003 for
details).

In order to estimate the $\gamma$-ray luminosity that could be
produced in the jet-wind interaction of a misaligned MQ with a high-mass companion,
we have fixed the parameters that describe the geometry of
this scenario for a standard case (Table \ref{t1}). Note that the
radius of the star holds a $\sim 26^\circ.5$ angle as seen from
the compact object, so the star fills $\sim 10.5\%$ of the compact
object's sky. We have calculated the luminosity in high-energy
photons for different jet-disk coupling values ($q_{\rm j}$), and for
different values of $B_\star$ and
 $B_{\rm j}(d_0)$. The most effective particle acceleration occurs at the relativistic plasma. We have considered a sub-relativistic strong shock that leads to a spectrum with $\Gamma\sim 2$ and a mildly relativistic shock with $u=0.5 c$ ($\Gamma\sim1.5$). Table \ref{tab2} characterizes (in the first three columns) a number of specific models, ordered by decreasing jet power, and then gives the location of the shock front, the maximum and minimum energy for protons penetrating into the base of the wind, and the integrated gamma-ray luminosities generated for the sub-relativistic and the relativistic cases of particle acceleration.

\begin{table*} %[t]
\caption{Some results about the high-energy photon emission
in misaligned MQs. The A, B, C letters naming the model indicate a
high, middle or low jet-disk coupling, respectively.}
\begin{center}
$\matrix{\hline \cr {\rm Model}&q_{\rm j} & B_{\star} & B_{\rm
j}(d_0) & r_{\rm s} & E_p^{\rm min} & E_{p,\,{\rm sub}}^{\rm max}&E_{p,\,{\rm rel}}^{\rm max} &L_{\gamma,(\Gamma=2)}&
 L_{\gamma,(\Gamma=1.5)}\cr
 &&[{\rm G}]&[{\rm
G}]&[R_\odot]&[{\rm TeV}]&[{\rm TeV}]&[{\rm TeV}]& [{\rm erg\,s^{-1}}]&[{\rm
erg\,s^{-1}}] \cr
\noalign{\smallskip}\hline \noalign{\smallskip} 
A1& 10^{-2}& 500 &10 & 21.7 & 92.2 & 190& 3000 & 2.2\,{{10}^{32}} & 4.1\,{{10}^{32}} \cr
A2& 10^{-2}& 500 &100 &21.7 &92.2 & 1900&30000& 7.5\,{{10}^{32}}  & 3.9\,{{10}^{32}} \cr B1& 10^{-3}& 1 & 10 & 34.2 & 0.2 & 190  & 3000 & 1.9\,{{10}^{32}} & 5.4\,{{10}^{31}} \cr
B2& 10^{-3}& 1 & 100& 34.2 & 0.2 & 1900 &30000& 2.1\,{{10}^{32}}  & 4.3\,{{10}^{31}}\cr
B3& 10^{-3}& 100 & 10 & 34.2 & 21.3& 190& 3000& 6.4\,{{10}^{31}}  & 4.7\,{{10}^{31}}\cr
B4& 10^{-3}& 100 &100 & 34.2 & 21.3& 1900&30000& 1.1\,{{10}^{32}} & 4.1\,{{10}^{31}} \cr C1& 10^{-4}& 1 & 10  &41.1 & 0.2 & 190 & 3000 & 1.9\,{{10}^{31}} & 5.4\,{{10}^{30}} \cr 
C2& 10^{-4}& 1 & 100 & 41.1& 0.2& 1900 & 30000& 2.1\,{{10}^{31}} & 4.3\,{{10}^{30}} \cr C3& 10^{-4}& 100 & 10 & 41.1& 20.8 &190& 3000& 6.5\,{{10}^{30}}  & 4.7\,{{10}^{30}} \cr
C4&10^{-4} & 100 & 100& 41.1& 20.8 &1900&30000& 1.1\,{{10}^{31}} &4.1\,{{10}^{30}} \cr
\noalign{\smallskip} \hline \cr}
   $
\label{tab2}
\end{center}
\end{table*}

\begin{figure*}%[h]
\centering \psfig{figure=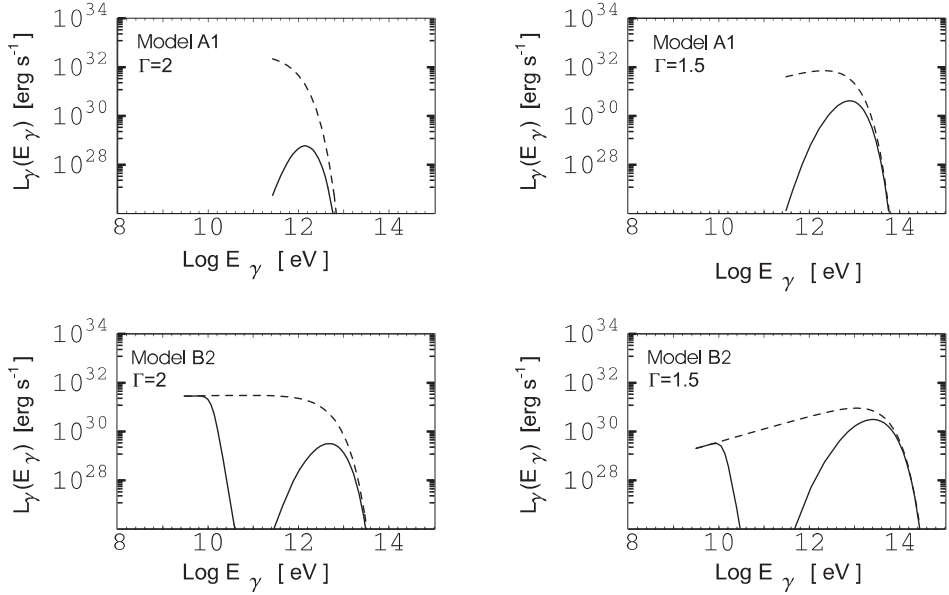, width=12.5cm}
\caption{Spectral high-energy distribution at TeV energies from
the jet-wind interaction in a misaligned MQ. The left panel
corresponds to $\Gamma=2$, and the right to $\Gamma=1.5$ (see
text). The dashed lines correspond to the production spectra whereas the solid lines represent the absorbed spectra.}\label{spectro}
\end{figure*}

The spectral high-energy distributions for some of the computed
models are shown in Figure \ref{spectro}. The plots correspond to
$L_\gamma(E_\gamma)=E_\gamma^2I_\gamma(E_\gamma)$, with an
exponential-type cutoff at $E_{\gamma}\sim ({E}_p^{\rm max}/{\rm
GeV})^{3/4} /20 $. The obtained isotropic luminosities above 1 TeV
are in the range $10^{30}-10^{32}$ erg s$^{-1}$.

\begin{figure*}%[h]
\centering \psfig{figure=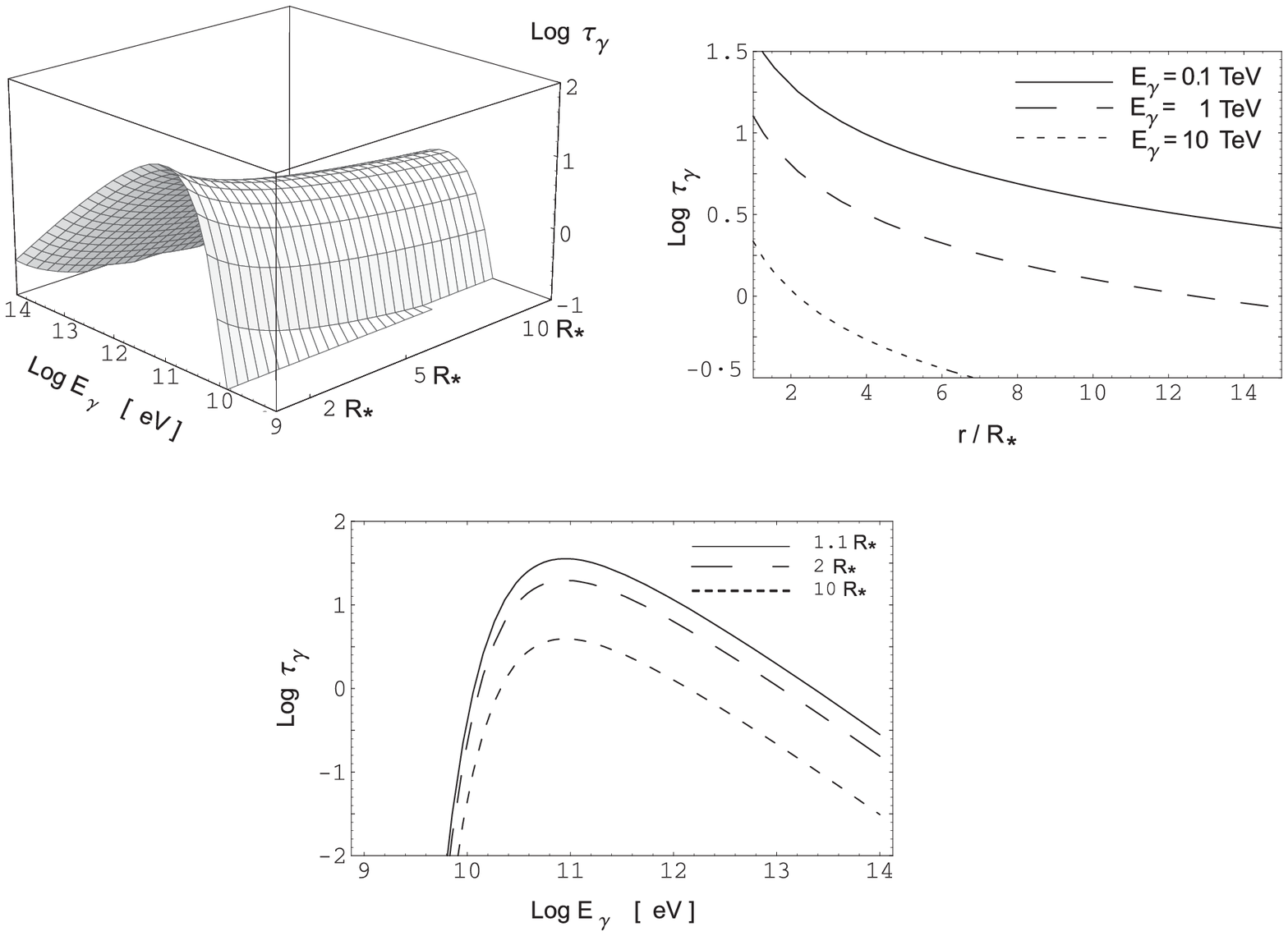, width=16.0cm} \caption{Stellar photosphere's opacity due to photo-pair creation, as function of
the $\gamma$-ray energy and the distance from the star, for a photon moving radially from the massive star.  Upper-right: Optical depth for fixed values of
energy. Lower panel: Optical depth for fixed distances.} \label{fig1}
\end{figure*}

High-energy gamma rays can be absorbed in the stellar photon field
through pair production, and then inverse Compton emission from
these pairs can initiate an $e^{\pm}$-pair cascade. This effect
has been studied in detail by Bednarek (1997, 2000) and more recently by
Sierpowska \& Bednarek (2005). Figure \ref{fig1} shows the
$\tau_\gamma$ opacity of the stellar photosphere to gamma-ray
propagation as a function of the gamma-photon energy and the
distance $r$ from the star where it is created. This opacity,
 for a photon moving radially from the massive star, can be roughly 
computed, averaging over the injection angles, as:
\be
\tau_\gamma(E_\gamma,r)=\int_r^\infty \int_{0}^\infty
N(E_\star,\,r') \sigma_{e^-e^+}(E_\star,E_\gamma) dE_\star dr',
\ee
where $E_\star$ is the energy of the photons emitted by the
star, $E_\gamma$ is the energy of the $\gamma$-ray, and
$\sigma_{e^-e^+}(E_\star,E_\gamma)$ is the cross section for
photon-photon pair production (e.g. Lang 1980). The stellar photon
distribution is that of a blackbody peaking at the star's
effective temperature ($T_{\rm eff}$):
\be
N(E_\star,\,r)=(\pi
B(E_\star)/h E_\star c) R_\star^2/r^2, \ee
where $h$ is the Planck
constant and
\be
B(E_\star)=[2{E_\star}^3/(hc)^2]/(e^{E_\star/kT_{\rm
eff}} -1).
\ee

The inverse Compton opacity $\tau_{\rm IC}(E_e,\,r)$ determines
the advance of the cascades after the pair creation. We have
estimated that gamma-rays with energies below $\sim 10^{12}$ eV
will be mostly absorbed, triggering IC cascades. More energetic
photons ($>10$ TeV) will escape. The result will be a soft
spectrum at GeV energies (Bednarek \& Protheroe 1997) and a peak at 
high energies (Figure \ref{spectro}).

%\begin{figure*}%[h]
%\centering \psfig{figure=tauIC.eps, width=8.0cm} \caption{Inverse Compton opacity of %the stellar photosphere to electron(positron)
%propagation. $\tau_{\rm IC}$ is a function of the $e^{\pm}$ energy
%and the distance from the star. Lower panel:
%Optical depth for fixed distances.} \label{tauIC}
%\end{figure*}

%\begin{figure}%[h]
%\centering \psfig{figure=tau-gamma-3Ebis.eps, width=8.0cm}
%\caption{\it Photosphere's opacity for fixed energies of 1 erg (solid line), 10 erg and 100 erg (dashed, higher and lower respectively).}\label{fig2}
%\end{figure}

\section{Neutrino production and possible detection}

The $\nu_\mu + \bar\nu_\mu$ neutrino intensity $I_\nu(E_\nu )$ may
be derived from the spectral $\gamma$-ray intensity by imposing
energy conservation (Alvarez-Mu\~niz \& Halzen 2002, Torres et al. 2004):
%(see also Stecker 1979):
\be \int E_\gamma I_\gamma(E_\gamma) dE_\gamma = C \int E_\nu I_\nu(E_\nu )
dE_\nu, \ee
where the limits of the integrals are ${E_{\gamma\;
[\nu]}^{\rm min}}$ ($E_{\gamma \; [\nu]}^{\rm max}$), the minimum
(maximum) energy of the photons (neutrinos) and $C$
is a numerical constant of order one (i.e. the total neutrino
luminosity produced by the decay of charged pions is a comparable
quantity with that of the photons $L_\gamma$). The neutrino
intensity will be given by
\be
I_\nu (E_\nu)=C_{\nu} (E_\nu /{\rm erg})^{-\Gamma}\, e^{-E_\nu/E_\nu^{\rm max}}, \ee 

where the constant $C_{\nu}$ is in erg$^{-1}$ s$^{-1}$. The values of
$C_{\nu,(\Gamma=1.5)}$ and $C_{\nu,(\Gamma=2)}$ for the different
models are listed in Table \ref{tab3}.

The neutrinos are produced in the inner wind around the star. A
fraction of them, which will depend on the specific geometry of
the system, will be intercepted by the star. The star, in turn,
can absorb some of these neutrinos. We have estimated the
neutrino energy deposition rate $\eta_{\nu}$ (see Gaiser et al. 1986) inside the star.
The results confirm that no effects of neutrino heating will be
observable during the stellar lifetime. This fact is expected since
only a small fraction of the power of the jet is directed toward the
star by neutrinos ($\eta_\nu\sim 10^{27}\,-\,10^{30}$ erg
s$^{-1}$) and this represents a very small fraction ($10^{-11}\,-\,
10^{-8}$) of the luminosity of the star.

Although the neutrino emission has no effect upon the companion
star, perhaps it could be detectable at the Earth. If the distance
to the system is $ d_{\rm MQ}$, the neutrino flux arriving to a
km-scale detector like ICECUBE will be
$F_\nu(E_\nu)=I_\nu(E_\nu)/(4\pi\,d_{\rm MQ}^2)$. To make some
quantitative estimate we adopt in what follows $d_{\rm MQ}\sim 5$
kpc.

The neutrino signal can be estimated as (Anchordoqui et al. 2003)
\be
S =T_{\rm obs}  \int dE_\nu A_{\rm eff} F_\nu(E_\nu) P_{\nu \to \mu}
(E_\nu), \ee where $P_{\nu
\to \mu} (E_\nu) \approx 1.3 \times 10^{-6} \,(E_\nu/{\rm
TeV)}^{0.8}$ denotes the probability that a $\nu$ of energy $E_\nu
\sim 1 - 10^3~{\rm TeV}$, on a trajectory through the detector,
produces a muon~(Gaisser et al. 1995). $T_{\rm obs}$ is the
observing time and $A_{\rm eff}$ the effective area of the
detector. On the other hand, the noise will be given by
\be
N=\left[T_{\rm obs}
  \int dE_\nu  A_{\rm eff} F_{\rm B}(E_\nu) P_{\nu \to \mu} (E_\nu)
 \Delta \Omega
\right]^{1/2},\ee
where $\Delta \Omega$ is the solid angle of the
search bin ($\Delta \Omega_{1^\circ \times 1^\circ} \approx 3
\times 10^{-4}$~sr for an instrument like ICECUBE,  Karle 2002)
and
\be F_{\rm B}(E_\nu) \leq 0.2 \,(E_\nu/{\rm GeV})^{-3.21}\;\; \rm{GeV}^{-1}\;
\rm{cm}^{-2}\; \rm{s}^{-1}\; \rm{sr}^{-1}\ee
is the $\nu_\mu + \bar \nu_\mu$
atmospheric $\nu$-flux~(Volkova 1980, Lipari 1993).
Using as the integration limits the minimum and maximum
energies detectable by the instrument (i.e. 1-10$^5$ TeV)
%for the neutrinos produced in each case ($E_\nu^{\rm
%max,\, min}\sim E_p^{\rm max,\, min}/12$)
we obtain $T_{\rm obs}$ for a signal-to-noise ratio $S/N= 3$. The
results are in all the cases (models A1 to C4) greater than several
thousands years,
%depending on the model, range from 300 yr (model $A1$) to $10^7$ yr (model $C4$)
 rendering the detection impractical with the current sensitivity. Once again, these are lower limits, since the duty cycle is not 1. Only in the case of very powerful jets with $q_{\rm j}>0.1$ a detection might be possible on human-life timescales. We will discuss this case in the next section.

\begin{table} %[t]
\begin{center}
\caption{Some results about the neutrino emission in misaligned
MQs. Case $D1$ corresponds to the direct impact of a jet with
$q_{\rm j}=0.1$ onto the stellar surface.}

$
\matrix{\hline \cr {\rm Model}& C_{\nu,(\Gamma=2)}&
C_{\nu,(\Gamma=1.5)}\cr
&{\rm erg^{-1}s^{-1}}&{\rm erg^{-1}s^{-1}}\cr
 \noalign{\smallskip} \hline \noalign{\smallskip}
A1& 4.0\,{{10}^{32}} & 1.7\,{{10}^{32}} \cr
A2& 3.3\,{{10}^{32}} & 5.4\,{{10}^{31}} \cr
B1& 3.7\,{{10}^{31}} & 1.6\,{{10}^{31}} \cr
B2& 3.1\,{{10}^{31}} & 5.4\,{{10}^{30}} \cr
B3& 3.9\,{{10}^{31}} & 1.7\,{{10}^{31}} \cr
B4& 3.3\,{{10}^{31}} & 5.4\,{{10}^{30}} \cr
C1& 3.7\,{{10}^{30}} & 1.6\,{{10}^{30}} \cr 
C2& 3.1\,{{10}^{30}} & 5.4\,{{10}^{29}} \cr
C3& 3.9\,{{10}^{30}} & 1.7\,{{10}^{30}} \cr 
C4& 3.3\,{{10}^{30}} & 5.5\,{{10}^{29}} \cr \noalign{\smallskip}\hline
\noalign{\smallskip}\cr 
D1& 3.7\,{{10}^{32}} & 3.1\,{{10}^{34}} \cr \noalign{\smallskip}
\hline \cr}
$
\label{tab3}
\end{center}
\end{table}

\section{Powerful jets and direct interaction with the star}\label{heavy-jets}

We consider now a heavy jet directly impacting the star ($q_{\rm
j}=0.1$\footnote{$L_{\rm j}=q_{\rm j}\,5.7\, 10^{38}$ erg
s$^{-1}$}). The proton spectrum of the relativistic jet flow is
assumed to be a power law $N'_{p}(E'_{p})= K_{\Gamma}\;
{E'}_{p}^{-\Gamma}$, valid for $ {E'_{p}}^{\rm min}\leq E'_{p}
\leq  {E'_{p}}^{\rm max}$, in the co-moving jet frame. The
corresponding particle flux will be $J'_{p}( E'_{p})= (c/4\pi)
N'_{p}(E'_{p})$. Since the jet expands (in a conical way), the
proton flux can be written as:
\begin{equation}
J'_p(E'_p)=\frac{c}{4 \pi} K_{\Gamma}
\left(\frac{d_0}{d}\right)^{n} {E'_p}^{-\Gamma},
\label{Jp}
\end{equation}
where $n$ is positive and we assume $n=2$ (which corresponds to
the conservation of the number of particles, see Ghisellini et al.
1985), and a prime refers to the jet frame. Using relativistic
invariants, it can be proved that the proton flux reaching the
surface of the star, in the lab frame, becomes (e.g. Romero et al.
2003)

\begin{eqnarray}
J_p(E_p)&=&\frac{c K_{\Gamma}}{4 \pi} \left(\frac{d_0}{a-R_\star}\right)^
{n} \nonumber \\ && \times   \frac{\gamma_{\rm bulk}^{-\Gamma} \left(E_p+\beta_{\rm b}
\sqrt{E_p^2-m_p^2c^4} \right)^{-\Gamma}}{\left(1 + \frac{\beta_{\rm b}
\;E_p}{\sqrt{E_p^2-m_p^2 c^4}}\right)}, \label{Jp_lab}
\end{eqnarray}
where $\gamma_{\rm bulk}$ is the jet Lorentz factor, $\beta_{\rm b}$
is the corresponding velocity in units of $c$, and we have considered protons with small transverse momentum (hence the lack of dependency in the particle angle when compared with the corresponding expression in Romero et al. 2003).

The number density ${n_0}'$ of particles
flowing in the jet at $R_0=R(d_0)$ can be determined as in Romero et al. (2003) and the normalization constant $K_{\Gamma}$ can be derived provided  that ${E'}_p^{\rm max}>>{E'}_p^{\rm min}$. Since ${E'}_p^{\rm min}\sim$ 1 GeV, this is always the case, and
\begin{equation}
K_{\Gamma}={n_0}' (\Gamma-1) ({E'}_p^{\rm min})^{\Gamma-1}.
\end{equation}
In the numerical calculations we have considered ${E'}_p^{\rm max}=100$ TeV, %${E'}_p^{\rm min}=10$ GeV,
$\gamma_{\rm bulk}=10$, and two cases for the proton energy distribution in the jet\footnote{Notice that these are the original particle distributions in the jet and not the result of re-acceleration at the colliding shock region as before. Here the jet impacts directly in a high-density region.}: $\Gamma=2$ and $\Gamma=1.5$.

The gamma-rays generated by the jet-star interaction will be absorbed by the electromagnetic cascades in the star. The main signature of such cascades could be a rather broad $e^+-e^-$ annihilation line. The neutrino flux from charged pion decays will be significantly larger than in the case of the jet-wind interactions.  %$L_{\gamma\,(\Gamma=2)}\simeq 2.7\,10^{33}$ erg s$^{-1}$, %$L_{\gamma\,(\Gamma=1.5)}\simeq 6.9\,10^{34}$ erg s$^{-1}$}
 For the density profile of the star we have used
\begin{equation}
\rho(r)=\rho_{\rm c}\left[1-(r/R_\star)^2\right],
\end{equation}
with $\rho_{\rm c}\simeq 0.05$ g/cm$^3$, in accordance with the adopted stellar mass.

In the lab frame, the neutrino energies reach $\sim 1$ TeV, then only the hard energy tail of the neutrino spectrum lies in the range of observation of ICECUBE. The
neutrino energy deposition rate for $\Gamma=1.5$ results $\eta_\nu
\simeq 2.4\, 10^{33}$ erg s$^{-1}$.
%, thus $t_B\sim 2\,\, 10^7$ yr.
%We see, then, that in this case the stellar evolution might be
%marginally affected by the neutrino irradiation. 
 The neutrino
signal on the Earth produces a number of muons per year of $N_{\mu}=3.25$. This means a detectable signal at $3\sigma$-level for an observing time $T_{\rm obs}\sim 15$ years, in the case of a source at 2 kpc and a close alignment of the jet with the line of sight and a duty cycle of 20\%.

%The estimates for neutrino emission and absorption are given in
%Table \ref{tab3} (Model $D1$).

\section{Comments}

The type of misaligned MQ discussed in Section \ref{gamma} is a
variable gamma-ray source. For the masses and orbital radius given
in Table \ref{t1}, the orbital period is $\sim2.5$ days. Since the
star occupies $\sim10$ \% of the sky of the compact object and
there are two jets in the orbital plane, the expected duty cycle
is $\sim 20$ \%. The gamma-ray emission is ignited twice per
orbit, every $\sim 30$ hours. The source remains active for around
$\sim 6$ hours each time, since the diffusion timescale for the
accelerated particles is much shorter than this value: $t_{\rm d} (E_p^{\rm max})\la 29$ s and 
$t_{\rm d}(E^{\rm min}_p)\la 203$ s for the models considered here. All these numbers will change, of course, from one
case to another depending on the specific configuration, but in
all cases the result will be a variable source unless the
diffusion timescale is much longer than the interval between the
different jet-wind collisions.

Because of the isotropization of the particle flux at the shock front located at $r_{\rm s}$ the gamma-ray emission will not be as much dependent on the viewing angle as in the case of pure relativistic jet models (e.g. Romero et al 2003). Occultation events will occur only if the observer lies close to the orbital plane. Such events would be rare since they require an alignment among compact object, star and observer. The flux, additionally, would not be completely suppressed since the relativistic particles diffuse around the entire inner wind.

Since misaligned MQs can be TeV and (through electromagnetic cascades in the stellar photosphere) MeV-GeV sources, we could ask whether some of the unidentified gamma-ray sources detected by EGRET instrument might be associated with them. There are $\sim 20$ unidentified variable sources at low galactic latitudes in the Third EGRET Catalog (Hartman et al. 1999, Torres et al. 2001, Nolan et al. 2003). Bosch-Ramon et al. (2004) have shown that they form a distinct population with a distribution consistent with what is expected for high-mass MQs. The total number of high-mass X-ray binaries brighter than $10^{34}$ erg s$^{-1}$ at X-rays in the Galaxy is estimated in $\sim 380$ (Grimm et al. 2002). Only a small fraction of them will have the type of persistent jets necessary to produce detectable gamma-ray sources. Around 20 sources seems to be a reasonable estimate (Bosch-Ramon et al. 2004). If the original orientation angles of the jets are randomly distributed we could expect that in 10-20 \% of these systems the jet could be periodically directed towards the companion star (Butt et al. 2003). This means that 2-4 of the variable EGRET sources at low galactic latitudes might be associated with misaligned MQs. Hints of periodic variability could help to identify these sources. Their main footprints, however, would be a stronger TeV emission than what is expected in normal MQs. Hence, future observations with Cherenkov telescopes like HESS, MAGIC or Veritas might lead to the identification of these sources.

\section{Conclusions}

The jets of MQs can form a small angle respect to the orbital plane. In some high-mass X-ray binary systems the jet could be directed straight to the companion star. We have shown that in such a situation the wind of the star can balance the jet pressure in many cases. A jet-wind collision region is formed between the star and the compact object, with a contact discontinuity separating both shocked media. Relativistic particles injected by the jet in the strong shock can be re-accelerated up to very high energies and isotropized in this region. These particles diffuse into the inner wind as far as diffusion dominates over convection. Gamma-rays and neutrinos can be then produced in the densest part of the wind. We have presented calculations of the gamma-ray luminosity, which might be detectable through high-energy telescopes. The neutrino flux is usually below the sensitivity of km-scale detectors like ICECUBE unless extremely powerful jets are assumed. Such jets could reach the star, triggering cascades and producing nucleosynthesis there, as discussed by Butt et al. (2003). We have also shown that photon-photon absorption of gamma-rays with energies below 10 TeV can initiate electromagnetic cascades in the stellar photosphere, leading to MeV-GeV sources that might contribute to the population of variable gamma-ray sources detected by EGRET at low galactic latitudes. Future instruments like GLAST will help to unveil the nature of these sources.

\begin{acknowledgements}
We thank J.M. Paredes and V.Bosch-Ramon for valuable discussions and comments on the manuscript. We also thank an anonymous referee who made important suggestions that led to a significant improvement of the manuscript. 
This research has been supported by Fundaci\'on Antorchas, CONICET, and the
Argentine agency ANPCyT through Grant PICT 03-13291.

\end{acknowledgements}

{}


\begin{thebibliography}{}
\bibitem{}Aharonian, F. A., \& Atoyan, A. M. 1996, A\&A, 309, 917
\bibitem{}Anchordoqui, L.A., Torres, D.F., McCauley, T., et al. 2003, ApJ, 589, 481
\bibitem{}Alvarez-Mu\~niz,  J. \& Halzen, F. 2002, ApJ, 576, L33
\bibitem{}Bardeen, J.M., \& Petterson, J.A. 1975, ApJ, 195, L65
\bibitem{}Bednarek, W. 1997, A\&A, 322, 523
\bibitem{}Bednarek, W. 2000, A\&A, 362, 646
\bibitem{}Bednarek, W. \& Protheroe, R. J. 1997, MNRAS, 287, L9
\bibitem{}Biermann, P. L. \& Strittmatter, P. A. 1987, ApJ, 322, 643
\bibitem{}Blandford, R. D. \& Ostriker J. P. 1978, ApJ, 221, L29
\bibitem{}Bosch-Ramon V., Romero, G.E., \& Paredes J.M. 2005, A\&A, 429, 267
\bibitem{}Butt, M. Y., Maccarone, J. T. \& Prantzos, N. 2003, ApJ, 587, 748
\bibitem{}Conti, P. S. \& Ebbets, D. 1977 ApJ, 213, 438
%\bibitem{}Dermer, C. D. 1986, A\&A, 157, 223
\bibitem{}Donati, J.F. et al. 2002, MNRAS, 333, 55
%\bibitem{}Drury, L.O'C., Aharonian, F.A., \& V\"olk, H.J. 1994, A\&A, 287, 959
\bibitem{}Falcke, H. \& Biermann, P. L. 1995, A\&A, 293, 665
\bibitem{}Gaisser, T. K., Stecker, F.W., Harding A.K., \& Barnard, J.J. 1986, ApJ, 309, 674
\bibitem{Gaisser:1994yf}
Gaisser, T.K., Halzen, F., \& Stanev, T. 1995,
%``Particle astrophysics with high-energy neutrinos,''
Phys.\ Rept., 258, 173 [Erratum-ibid., 271, 355 (1996)]
\bibitem{}Gaisser, T.K. 1990, Cosmic Rays and Particle Physics, Cambridge
University Press, Cambridge
\bibitem{}Ghisellini, G., Maraschi, L., \& Treves, A. 1985, A\&A, 146, 204
\bibitem{} Ginzburg, V. L. \& Syrovatskii S. I. 1964, The Origin of Cosmic Rays, Pergamon Press, New York
\bibitem{}Grimm, H.-J., Gilfanov, M., Sunyaev, R. 2002, A\&A, 391, 923
\bibitem[1999]{Hartman99}Hartman, R. C., Bertsch, D.~L., \& Bloom, S.~D.~et~al. 1999, \apjs, 123, 79
\bibitem{}Katz, J.I. 1980, ApJ, 236, L127
\bibitem{}Kaufman Bernad\'o, M. M., Romero, G. E., \& Mirabel, I. F. 2002, A\&A, 385, L10
\bibitem{}Kopal Z., 1959, Close binary systems, J.Wiley \& Sons, New York
\bibitem{}Kotani, T., Kawai, N., Aoki, T., et al. 1994, PASJ, 46, L147
\bibitem{}Kotani, T., Kawai, N., Matsuoka, M., \& Brinkmann, W. 1996, PASJ, 48, 619
%\bibitem{}Lamers, H. \& Leitherer, C. 1993, ApJ, 412, 771
\bibitem{}Lamers, H.J.G.L.M. \& Cassinelli, J.P. 1999, Introduction
to Stellar Winds,  Cambridge University Press, Cambridge
\bibitem{}Lang, K. R. 1980, Astrophysical Formulae, Springer-Verlag, Berlin
\bibitem{}Larwood, J. 1998, MNRAS, 299, L32
\bibitem{}Lipari, P. 1993,
%``Lepton Spectra In The Earth's Atmosphere,''
Astropart.\ Phys.\,  1,  195
%%CITATION = APHYE,1,195;%%
\bibitem{}Maccarone, J.T. 2002, MNRAS, 336, 1371
\bibitem{}Mannheim, K. \& Schlickeiser, R. 1992, A\&A, 286, 983
\bibitem{}Massi M., Rib\'o J.M., Paredes J.M, et al. 2004, A\&A, 414, L1
\bibitem{}Migliari, S., Fender, R. \& M\'endez, M. 2002, Science, 297, 1673
\bibitem{}Mioduszewski, A.J., et al. 2001, ApJ, 553, 766
\bibitem{}Mirabel, I. F., \& Rodr\'{\i}guez, L. F. 1999, ARA\&A, 37, 409
\bibitem[2003]{Nolan03}Nolan, P.L, Tompkins, W.F., Grenier, I.A., \& Michelson, P.F. 2003, ApJ, 597, 615
\bibitem{}Protheroe, R. J. 1998, ADP-AT-98-9 [astro-ph/9812055]
\bibitem{}Ogilvie, G.I., \& Dubus, G. 2001, MNRAS, 320, 485
\bibitem[1999]{Romero99} Romero, G.~E., Benaglia, P., \& Torres, D.~F. 1999, A\&A, 348, 868
\bibitem{}Romero, G.E., Kaufman-Bernad\'o, M.M. \& Mirabel, I.F. 2002, A\&A 393, L61
\bibitem{}Romero, G.E., \& Torres, D.F. 2003, ApJ, 586, L33
\bibitem{}Romero, G.E., Torres, D.F., Kaufman-Bernad\'o, M.M. \& Mirabel, I.F. 2003,
A\&A, 410, L1
\bibitem{}Romero, G.E. 2004, Chin. J. A\&A, in press [astro-ph/0407461]
\bibitem{}Sarazin, C.L., Begelman, M.C., \& Hatchett, S.P. 1980, ApJ, 238, L129
\bibitem{}Sierpowska, A. \& Bednarek, W. 2005, MNRAS 356, 711
\bibitem{}Stecker, F. W. 1979, ApJ, 228, 919
\bibitem[2001]{Torres01}Torres, D.~F., Romero, G.~E., Combi, J.~A., Benaglia, P., Andernach, H., \& Punsly, B.\ 2001,
\aap, 370, 468
\bibitem{}Torres, D.F., Domingo-Santamar\'{\i}a E., \& Romero, G.E. 2004, ApJ, 601,L75
%\bibitem{}Torres, D.F., et al. 2003, Physics Reports, in press;
%astro-ph/0209565
\bibitem{}Vacca, W.D., Garmany, C.D., \& Shull, J.M. 1996 ApJ 460, 914
\bibitem{} V\"olk, H.J., \& Forman, M. 1982 ApJ 253, ~188
\bibitem{Volkova:sw}
Volkova, L.V. 1980,
%``Energy Spectra And Angular Distributions Of Atmospheric Neutrinos,''
Sov.\ J.\ Nucl.\ Phys.\, 31, 784
\bibitem{} Weber, E.J., \& Davis, L. 1967, ApJ, 148, 217
\bibitem{} White, R.L. 1985, ApJ, 289, 698
\bibitem{} Wijers, R.A.M.J., \& Pringle, J.E. 1999, MNRAS, 308, 207

\end{thebibliography}
\end{document}